\documentclass[aps,prb,twocolumn,amsmath,amssymb,showpacs,prl,unsortedaddress]{revtex4-1} 
\usepackage{epsf}      
\usepackage{graphicx}
\usepackage{gensymb}
\usepackage{sidecap}

\begin{document}

\title {Dramatic Changes in the Electronic Structure Upon Transition to the Collapsed Tetragonal Phase in CaFe$_2$As$_2$}

\author{R. S. Dhaka}
\author{Rui Jiang}
\author{S. Ran}
\author{S. L. Bud'ko}
\author{P. C. Canfield}
\affiliation{Division of Materials Science and Engineering, The Ames Laboratory, U.S. DOE}
\affiliation{Department of Physics and Astronomy, Iowa State University, Ames, Iowa 50011, USA}
\author{Milan Tomi\`c}
\author{Roser Valent\'{\i}}
\affiliation{Institut f\"ur Theoretische Physik, Goethe-Universit\"at Frankfurt, Max-von-Laue-Stra\ss e 1, 60438 Frankfurt am Main, Germany}
\author{Yongbin~Lee}
\affiliation{Division of Materials Science and Engineering, The Ames Laboratory, U.S. DOE}
\author{B.~N.~Harmon}
\author{Adam Kaminski}
\affiliation{Division of Materials Science and Engineering, The Ames Laboratory, U.S. DOE}
\affiliation{Department of Physics and Astronomy, Iowa State University, Ames, Iowa 50011, USA}

\date{\today}                                         

\begin{abstract}
We use angle-resolved photoemission spectroscopy (ARPES) and density functional theory (DFT) calculations to study the electronic structure of CaFe$_2$As$_2$ in previously unexplored collapsed tetragonal (CT) phase. This unusual phase of the iron arsenic high temperature superconductors was hard to measure as it exists only under pressure. By inducing internal strain, via the post growth, thermal treatment of the single crystals, we were able to stabilize the CT phase at ambient-pressure.  We find significant differences in the Fermi surface topology and band dispersion data from the more common orthorhombic-antiferromagnetic or tetragonal-paramagnetic phases, consistent with electronic structure calculations.
The top of the hole bands sinks below the Fermi level, which destroys the nesting present in parent phases. The absence of nesting in this phase along with apparent loss of Fe magnetic moment, are now clearly experimentally correlated with the lack of superconductivity in this phase.
\end{abstract}

\pacs{74.25.Jb,74.62.Dh,74.70.-b,79.60.-i}
\maketitle

The AFe$_2$As$_2$ (A = Ca, Sr, Ba) materials have become one of the key systems for the study of Fe-based high temperature superconductivity \cite{Kamihara08,Takahashi08,XHChen08,TYChen08,GFChen08,Rotter08,Cruz08,SefatPRL08, Ni08, Canfield09}.  Detailed substitution and pressure studies on these systems have revealed that superconductivity is intimately linked to  the magnetic state of iron and can be turned on or off by manipulating the nature of the order, the fluctuations or even the existence of Fe-magnetism \cite{Rotter08, Ni08,Cruz08,SefatPRL08,Canfield09,Johnston10,CanfieldRev10,Hirschfeld,LeePRB09,AlirezaJPCM08}.  Moreover, in contrast to cuprate superconductors, the Fe-based superconductors are multiorbital systems and orbital degrees of freedom have been also proposed to be important for the understanding of the structural, magnetic and superconducting properties of these materials \cite{Krueger, Lee1, Lee2}.
%
At ambient pressure CaFe$_2$As$_2$ manifests a strongly first order, coupled structural / magnetic phase transition at 170 K that is exceptionally pressure sensitive with a remarkably large pressure derivative \cite{Ni08, Canfield09}. For pressures as low as 0.4 GPa another dramatic, first order phase transition to a non-moment bearing collapsed tetragonal phase \cite{Ni08} is stabilized near 100 K and rapidly increases in temperature for higher applied pressures. Whereas the  electronic properties of  CaFe$_2$As$_2$  in the ambient pressure phases \cite{LiuPRL09,KondoPRB10} were found to be similar to the other parent compounds of 122 family \cite{ShimojimaPRL10, YangPRL09, SatoPRL09, FinkPRB09, JensenPRB11, GuodongPRB09, ZhouPRB10,YiPRB09}, the electronic behavior of the CT phase is so far largely unexplored due to the need for at least 0.4 GPa of external pressure.

Post growth annealing/quenching of CaFe$_2$As$_2$ samples grown from excess FeAs can be used to control the degree of internal strain due to Fe-As precipitates associated with a small width of formation \cite{RanPRB11, RanPRB12, GatiPRB12}. 
%
 %
 One of the  important findings of these works is that the non-magnetic, collapsed tetragonal phase can be stabilized in an ambient pressure sample by using internal strain.  Key spectroscopic tools, such as ARPES and STM, which normally cannot be combined with pressures even at the 0.1~GPa level, can now be brought to play in order to understand the electronic properties of the collapsed tetragonal phase.

In this Letter we present ARPES measurements of the electronic properties of CaFe$_2$As$_2$ in the CT phase and compare them with the orthorhombic and (non-collapsed) tetragonal phases. Our data demonstrate that the band dispersion and FS topology change significantly with the transition to the CT phase, in agreement with our electronic structure calculations.  In particular the tops of the bands associated with Fe d$_{xy}$  and d$_{xz}$+d$_{yz}$ states at the center of the Brillouin zone move from above $E_{\rm F}$ to just below $E_{\rm F}$, eliminating the hole pockets.  Whereas there are significant changes elsewhere (to maintain electron count), loss of the nesting between the hole and electron sheets brought about by the shifts we observe are of key importance for magnetic and superconducting states in the low pressure phase \cite{DhakaPRL13}. 
Our findings provide  a peek into a phase that was previously accessible only under external pressure and provide further evidence of an intimate relation between the nature and nesting of the Fermi surface and magnetism in the iron arsenic high temperature superconductors.

\begin{figure}
\includegraphics[width=3.4in]{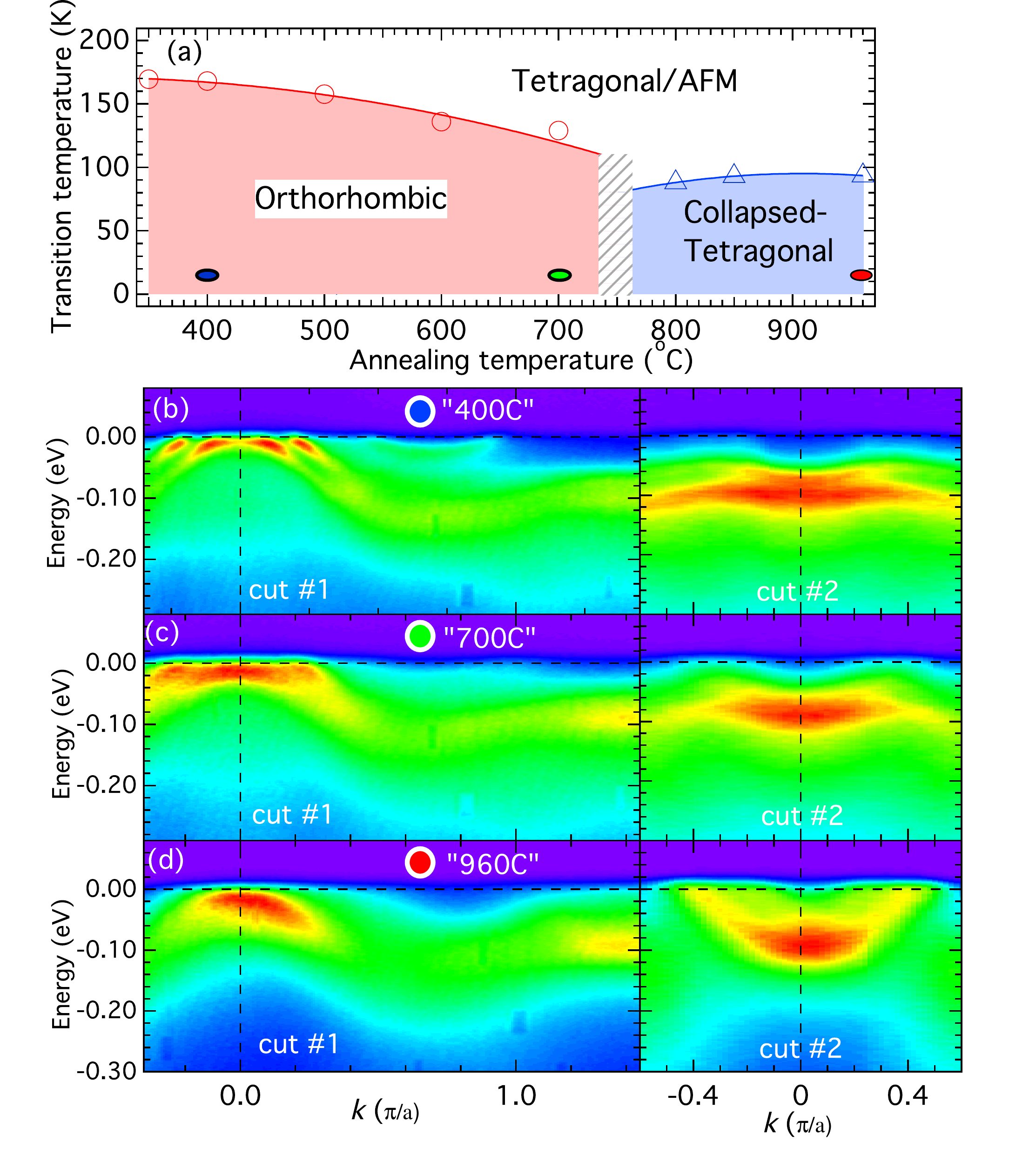}
\caption{(a) Schematic phase diagram for CaFe$_2$As$_2$ system based on data in Ref.~\cite{RanPRB11}. Blue, green and red circles mark the annealing condition of samples used in this study. The hatched area marks the boundary between two distinct phases. (b-d) band dispersion along cuts \#1 and \#2 for the three samples measured at T=15~K and $k_z=$15.3$\pi/c$.}
\label{fig1}
\end{figure}

\begin{figure*}
\includegraphics[width=6in]{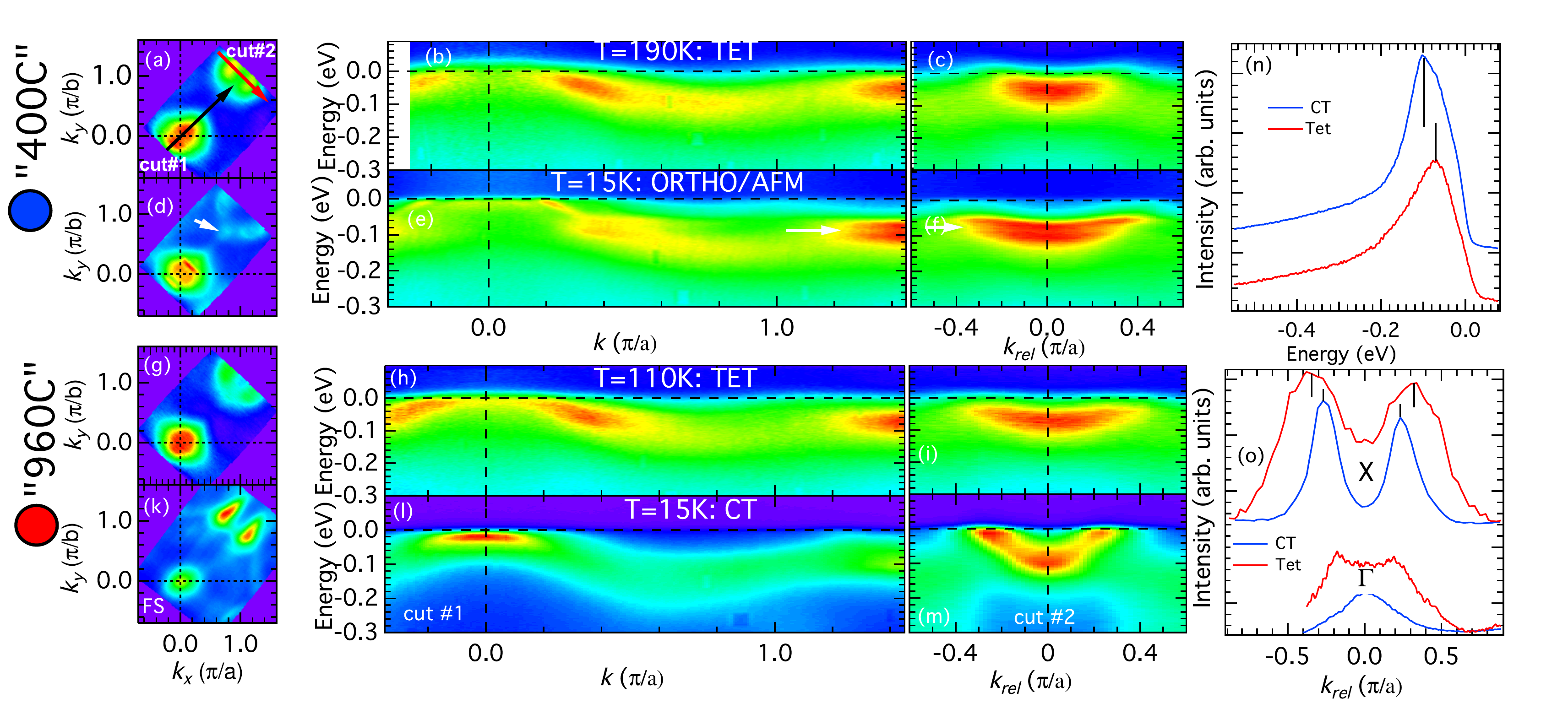}
\caption{The band structure of CaFe$_2$As$_2$ in three different phases measured at $k_z=$13.5 $\pi/c$. (a) Fermi surface map (b) band dispersion along cut \#1 (marked in panel a) (c) band dispersion along cut \#2 measured at 190~K in tetragonal/paramagnetic phase for a sample that was post-growth annealed at 400$^\circ$C. (d--f) the same plots for the same sample measured at 15K in orthorhombic/antiferromagnetic phase. Arrow in panel (d) points to small pockets associated with AFM order, while arrows in panels (e,f) point to band splitting caused by AFM order. (g--i) the same plots for a sample that was post-growth annealed at 960$^\circ$C measured at 110~K again in tetragonal/paramagnetic phase. (k--m) the same plots for the ``960C" sample, but measured at T=15~K in the CT phase. (n) EDCs at the corner of the zone, k=(1,1). The location of the peak marks the energy of the bottom of the electron pocket. (o) MDCs at the center of the zone - bottom two curves and corner of the zone - top two curves.}
\label{fig2}
\end{figure*}

\begin{figure}
\includegraphics[width=3.5in]{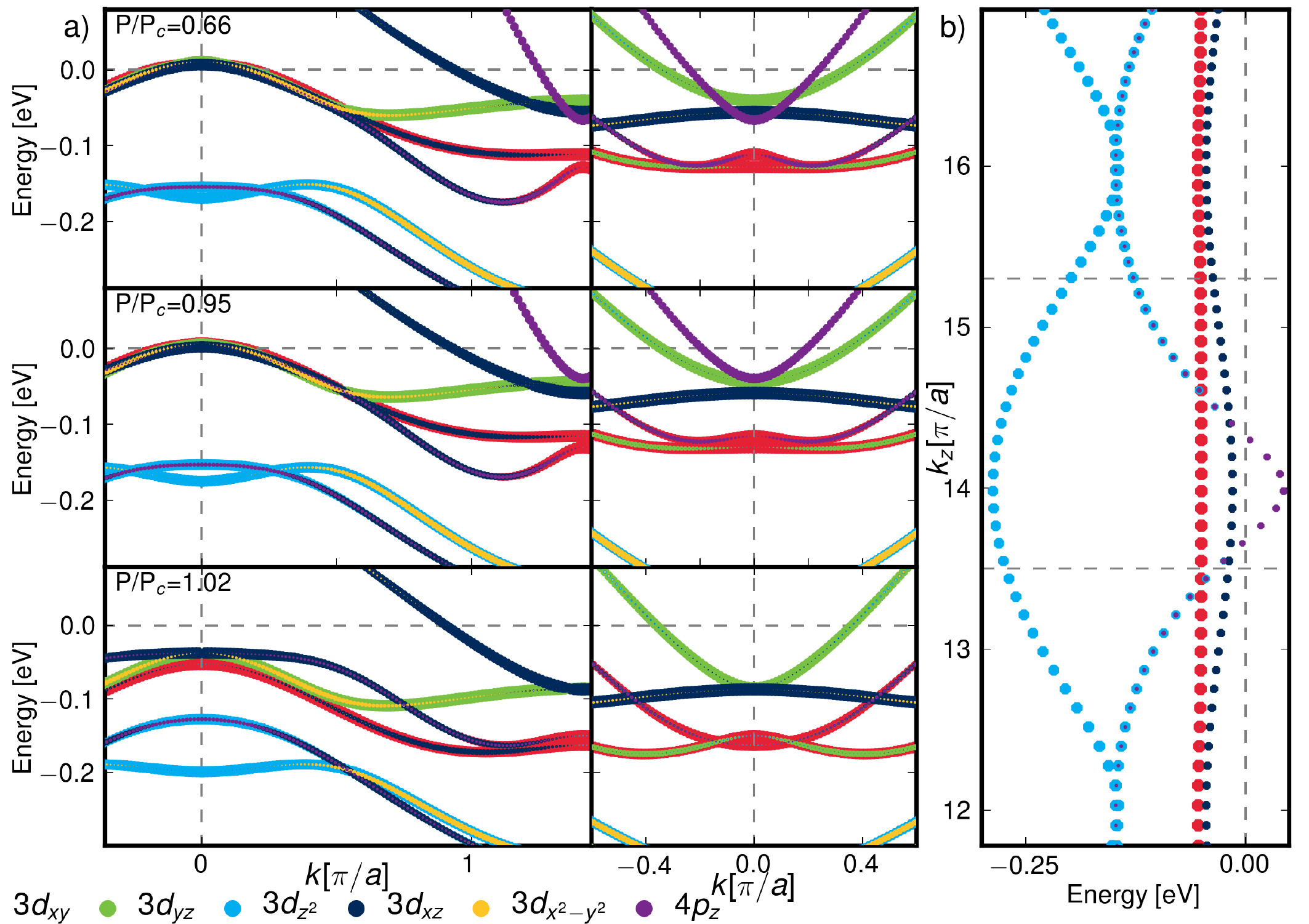}
\caption{Results of the DFT calculation of the band structure in CT at $k_z=$15.3$\pi/c$ corresponding to ARPES data shown on Fig. 1. (a) Left column shows band dispersion at the zone center, and right column corresponds to zone corner. (b) Calculated band dispersion along k$_z$ direction. Dashed lines mark the $k_z$ values of data in Figs. 1 and 2. Colors denote orbital contributions.}
\label{fig3}
\end{figure}

Single crystals of CaFe$_2$As$_2$ were grown from excess FeAs by slowly cooling a melt of CaFe$_4$As$_4$ from 1180$^\circ$C to 960$^\circ$C and then decanting off the excess liquid, essentially quenching the samples from 960$^\circ$C to room temperature, which is referred as ``960C" sample \cite{RanPRB11}. Upon cooling this sample becomes  nonmagnetic and has CT unit cell with significantly reduced $c$/$a$ ratio for T$\lesssim$100~K. This is remarkably similar to what has been observed for Sn grown CaFe$_2$As$_2$ samples under hydrostatic pressure (0.6~GPa)\cite{RanPRB11}. Post growth thermal treatments of these samples involve annealing at temperatures ranging from 350 to 800$^\circ$C for 24 hours, and subsequently quenching them from this annealing temperature to room temperature. Here, we have measured three samples identified by their annealing/quenching temperatures:  ``400C", ``700C"  and ``960C". As shown in Fig. 1a, these three samples cover the salient parts of the phase diagram: $\sim$170~K transition to low-temperature AF/orthorhombic (``400C"); $\sim$120K transition to AF/orthorhombic  (``700C"), and $\sim$100~K transition to non-magnetic CT phase (``960C"). At ambient conditions, the properties of the last two samples are remarkably similar to ones where hydrostatic pressure is applied to Sn grown samples of 0.35 and 0.675 GPa \cite{Canfield09, RanPRB11}. ARPES measurements have been performed (in grazing incidence geometry shown in Fig.~1(v) of Ref.~\onlinecite{DhakaPRL11}) at beamline 10.0.1 of the Advanced Light Source (ALS) using a Scienta R4000 electron analyzer. The measurements at Ames Laboratory were acquired using a Scienta SES2002 electron analyzer and a GammaData  helium ultraviolet lamp. The samples were mounted on an Al pin using UHV compatible epoxy and {\it in situ} cleaved  perpendicular to the $c$-axis, yielding single layer surfaces in the $a$-$b$ plane. All ARPES data were collected  in ultrahigh vacuum with pressures better than 4$\times$10$^{-11}$ torr. The energy and angular resolutions were set at $\approx15$~meV and $\approx0.3^{\circ}$, respectively. The Fermi level ($E_{\rm F}$) of the samples was referenced by measuring for a Au film evaporated {\it in-situ} onto the sample holder. High symmetry points were defined the same way as in Refs.~\onlinecite{LiuNP10}. The measurements carried out on several samples yielded similar results for the band dispersion and FSs. 
Ab-initio calculations~\cite{Tomic1,Tomic2} based on density functional theory as implemented in the Vienna ab initio simulations package (VASP)~\cite{Kresse} combined with the Fast Inertial Relaxation Engine (FIRE) algorithm~\cite{FIRE} were performed at finite pressure, both  for compressive and tensile stress. A collapsed tetragonal phase was obtained in the  calculations only for isotropically applied pressures as well as uniaxial pressures along the c-axis.

The evolution of the band structure with annealing/quenching temperature, which is equivalent to applying pressure, for  $k_z=15.3~\pi/c$ is shown in Figure 1. Because there is a significant decrease in the c-axis lattice parameter in the CT phase, we used different photon energies in order to compare data at similar values of $k_z$. This momentum plane was selected for  best signal to background ratio and clarity of the band dispersion in data (matrix elements). This momentum plane is accessed using 45~eV photons for orthorhombic/tetragonal phases and 62~eV for the CT phase. Panels b--d show the band dispersion along two cuts (marked in Fig. 2a)  for ``400C", ``700C" and ``960C" samples, respectively. The first two samples remain in orthorhombic/antiferromagnetic phase at low temperatures. At this value of $k_z$ and T=15~K, the matrix elements and relatively low scattering allow observation of two well separated hole bands in ``400C" sample. These bands broaden in  ``700C" sample, most likely due to some degree of inhomogenous stress being present. The ``960C" sample at low temperature is in the CT phase, which is  equivalent to application of 0.7 GPa of hydrostatic pressure to stress free, Sn flux grown samples\cite{Canfield09, RanPRB11, RanPRB12}. The band structure in this phase is very different indeed as illustrated in panel (d). The hole pockets sink below the Fermi energy while the electron band moves to higher binding energy and becomes more dispersive and sharper.  The vanishing of large portions of the hole Fermi sheets in CT phase resembles to a degree situation in its structural analog - CaFe$_2$P$_2$\cite{ColdeaPRL09}.

In Figure 2 we plot the Fermi surface and band dispersion in all three phases that occur in  the CaFe$_2$As$_2$. The tetragonal/paramagnetic phase can be accessed at higher temperatures in both ``400C" and ``960C" samples (see phase diagram in Fig. 1a). The data in orthorhombic and tetragonal phases was measured using 35~eV, which corresponds to  $k_z=13.5~\pi/c$. The same value of $k_z$ is accessed in CT phase when using 45~eV photons.  The Fermi surface maps for these phases are shown in panels (a) and (g) and the band dispersions along two symmetry cuts are shown in panels (b, c) and (h, i) for ``400C" and ``960C" samples, respectively. The electronic structure of the tetragonal phase consists of two hole pockets at the center of the zone and two electron pockets at the corners of the zone. Due to increased scattering at high temperatures the pairs of bands appear as a single, broader streak of intensity. These high temperature data are very similar to data measured in parent compounds in the 122 family at ambient pressures \cite{LiuPRL09, DhakaPRL11, VilmercatiPRB09}. The ``400C" sample has similar properties to samples at ambient pressure \cite{RanPRB11} e.~g. grown using Sn flux that are stress free (see phase diagram in panel 1a). Upon cooling the ``400C" sample enters the well documented orhtorhombic/antiferromagnetic phase. This is evident from data in Figs. 2(d,e,f), where complicated Fermi surface reconstruction (straight, nested sections of the hole pockets and sharp, high intensity points in electron pockets) and splitting of the bands at high energy is present. This is fully consistent with previous data in the AFM state \cite{LiuPRB09, YangPRL09, ZhangPRL09, KondoPRB10, LiuPRL09, YiPRB09} and signifies that ``400C" samples are indeed ambient/low pressure equivalent.   

\begin{figure}
\includegraphics[width=3.55in]{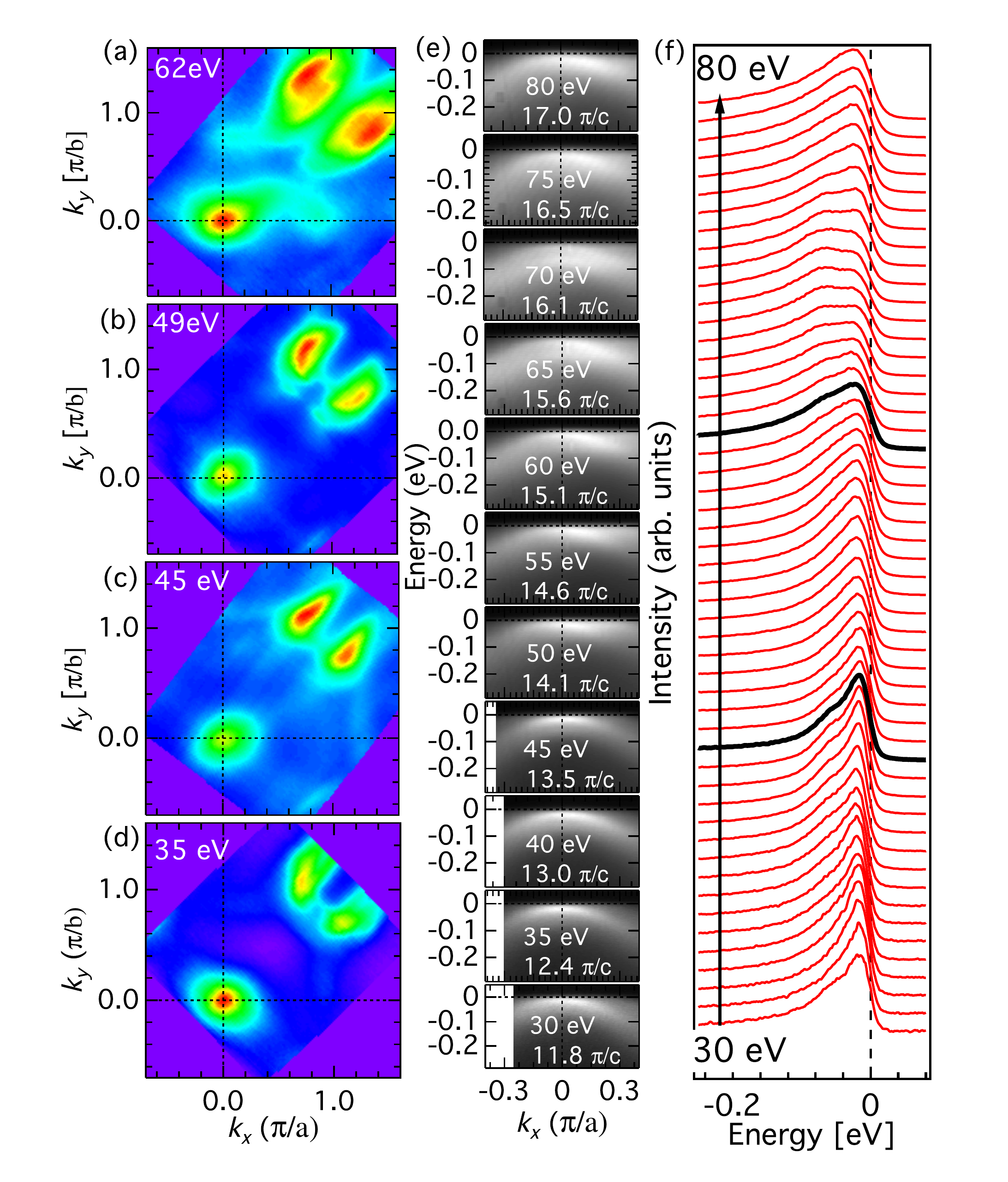}
\caption{FS maps of CaFe$_2$As$_2$ ``960C" sample measured with (a) 35, (b) 45, (c) 49 and (d) 62~eV photon energies at T$_{\rm S}\approx$15~K. (e) intensity plots close to the center of the zone along cut\#1 for photon energies from 30 eV to 80eV.  (f) EDCs at the center of the zone for photon energies from 30~eV to 80~eV, plotted every 1 eV. Thick (black) lines mark photon energies of 45 eV and 62 eV used in Figs. 1 and 2.}
\label{fig4}
\end{figure}

As previously mentioned,  the ``960C" sample shows a transition from the high temperature tetragonal phase (Fig. 2g-i) to the CT phase (Fig. 2 k-m) upon cooling.  Overall, the band dispersion in the CT phase at this $k_z$ plane (panels l and m) is very similar to data in Fig.~1. The appearance of brighter spots at different (from data in Fig. 1d) momentum points for both hole and electron bands  signifies modifications in the matrix elements due to different photon energies used. Overall the electronic properties in the CT phase agree qualitatively with the recently reported theoretical calculations  \cite{Tomic1, Tomic2, SannaPRB12}. Perhaps the most striking characteristic of the CT phase is the sinking of the hole pockets under the $E_{\rm F}$ and associated with this lack of the Fermi surface nesting. We quantify the key changes in the band dispersion by plotting the Energy Distribution Curves (EDCs) and Momentum Distribution Curves (MDC) at the high symmetry points in panels (n) and (o). In the CT phase, the bottom of the electron pocket moves to higher binding energy from $\sim$70 meV to $\sim$90 meV, as shown in panel (n). At the same time the diameter of the electron pocket shrinks slightly and the MDC peaks become sharper (top two curves in panel (o)). The MDC close to zone center have two peaks in the tetragonal phase, which signifies presence of a finite size hole pocket. Upon transition to the CT phase only weaker, single peak centered at the (0,0) is present (lower curve in panel (o)) that is consistent with an absence of the hole pocket at this $k_z$ plane and the top of this band being located below the chemical potential.

Results of DFT calculations of the band structure across the orthorhombic/CT phase transition are shown in Fig. 3.  We observe that simulations under hydrostatic pressure mimic ARPES data most closely, when renormalization by a factor of five is employed~\cite{correlations}. In particular, we find close correspondence between the observed Fermi surface behavior of "700C" and "960C" structures and P/Pc=0.95, and P/Pc=1.02 hydrostatic pressure structures, in the immediate vicinity of the Orthorhombic-to-CT phase transition at Pc.  At the transition, the d$_{xy}$, d$_{yz}$ and d$_{xz}$ bands abruptly drop below the Fermi level around the $\Gamma$-point, causing hole cylinders parallel to kz-axis to quickly disappear completely within the collapsed tetragonal phase, as observed in the ARPES measurements. These changes in the electronic structure show clearly that the phase transition tetragonal/CT and orthorhombic/CT  cannot be explained through a rigid band shift, as has been also discussed  in Co-doped Ca122 \cite{Damascelli}. In panel (b) we plot the calculated band dispersion in the CT phase along $z$-direction at the center of the zone.  Except for a very small crossing at $E_{\rm F}$ by one of the bands around the $k_z=14~\pi/c$, most of the bands are located below the chemical potential.

To examine the electronic properties along the $z$-direction and ensure that the differences we reported in the CT phase do not arise due to some unforeseen complications related to $k_z$ dispersion, we measured the FS and band structure using wide range of photon energies. The FS maps for selected photon energies are shown in Figs. 4 (a--d). There is very little variation in overall shape of the Fermi surface. The intensity of the inner and outer electron pockets changes with photon energy due to matrix elements, but their separation remains almost constant. 
Intensity plots along cut \#1 in CT phase measured using several photon energies are shown in Fig. 4e. The top of the band at the center of the zone appears to be consistently located below the chemical potential and the crossing predicted by the calculations (see Fig. 3b) is not visible in the data directly. This could be due to matrix elements, or overwhelming intensity from the broad bands located below $E_{\rm F}$. In Fig. 4f we plot EDC curves at the zone center as a function of photon energy with 1~eV step.  In addition to a sharp peak just below $E_{\rm F}$, a second peak at higher binding energy can be observed for range of $k_z$ values centered at $\sim$12.5~$\pi/c$ and $\sim$16~$\pi/c$. These structures most likely correspond to overlap of  d$_{xz}$ and d$_{z^2}$ bands shown in Fig. 3b. Based on DFT calculation, one would expect two broader, main peaks at binding energies of -50 meV and -150 meV, while those peaks are observed in the data at -18 meV and -68 meV. 
This discrepancy can be attributed to correlation effects which are not properly accounted for in DFT calculations. These effects can lead to strong band renormalizations \cite{Kotliar, Valenti}.The Fermi crossing around the $k_z=14~\pi/c$ is not directly seen in the data, most likely due to relatively weak intensity of this band.

In conclusion, our results reveal the electronic structure of  CT phase in unsubstituted CaFe$_2$As$_2$. Our data demonstrates that the FS and band dispersion in CT phase is significantly different from ambient pressure phases and in qualitative agreement with theoretical calculations for this phase. The hole pockets do not extend over appreciable range of the $k_z$ values and the destruction of nesting is accompanied by the vanishing of Fe magnetic moments and order. The resulting lack of magnetic fluctuations could be responsible for the absence of  superconductivity in this phase. 

We thank Sung-Kwan Mo for instrumentation support at the ALS. This work was supported by the U.S. Department of Energy (DOE), Office of Science, Basic Energy Sciences, Materials Science and Engineering Division.  Ames Laboratory is operated for the U.S. DOE by Iowa State University under contract \# DE-AC02-07CH11358 (sample growth, ARPES measurements and data analysis). MT and RV thank the Deutsche Forschungsgemeinschaft for funding through grant SPP 1458 (DFT calculations). The Advanced Light Source is supported by the Director, Office of Science, Office of Basic Energy Sciences, of the U.S. Department of Energy under Contract No. DE-AC02-05CH11231.

{\it Note added.}  After submission of this manuscript, we became aware 
of study electronic structure of CT phase in chemically substituted Ca(Fe$_{1-x}$Rh$_x$)$_2$As$_2$ by Tsubota {\it et al.} \cite{Tsubota}.


\begin{thebibliography}{99}
\bibitem{correlations}  The fact that bands have to be renormalized
by a factor of 5 indicates that the effects of correlations are
important in this system under various pressure conditions.
\end{thebibliography}
\end{document}